# Bias in Generative AI

**Mi Zhou**
UBC Sauder School of Business
mi.zhou@sauder.ubc.ca

**Vibhanshu Abhishek**
UCI Paul Merage School of Business
vibs@uci.edu

**Timothy Derdenger**
CMU Tepper School of Business
derdenge@andrew.cmu.edu

**Jaymo Kim**
CMU Tepper School of Business
jaymok@andrew.cmu.edu

**Kannan Srinivasan**
CMU Tepper School of Business
kannans@cmu.edu

**Abstract**

This study analyzed images generated by three popular generative artificial intelligence (AI) tools – Midjourney, Stable Diffusion, and DALL·E 2 – representing various occupations to investigate potential bias in AI generators. Our analysis revealed two overarching areas of concern in these AI generators, including (1) systematic gender and racial biases, and (2) subtle biases in facial expressions and appearances.

Firstly, we found that all three AI generators exhibited bias against women and African Americans. Moreover, we found that the evident gender and racial biases uncovered in our analysis were even more pronounced than the status quo when compared to labor force statistics or Google images, intensifying the harmful biases we are actively striving to rectify in today's society.

Secondly, our study uncovered more nuanced prejudices in the portrayal of emotions and appearances. For example, women were depicted as younger with more smiles and happiness, while men were depicted as older with more neutral expressions and anger, posing a risk that generative AI models may unintentionally depict women as more submissive and less competent than men. Such nuanced biases, by their less overt nature, might be more problematic as they can permeate perceptions unconsciously and may be more difficult to rectify.

Although the extent of bias varied depending on the model, the direction of bias remained consistent in both commercial and open-source AI generators. As these tools become commonplace, our study highlights the urgency to identify and mitigate various biases in generative AI, reinforcing the commitment to ensuring that AI technologies benefit all of humanity for a more inclusive future.

**Significance Statement**

Generative AI tools, like Midjourney, Stable Diffusion, and DALL·E, can inadvertently perpetuate and intensify societal biases related to gender, race, and emotional portrayals in their images. Our study uncovers not only the systematic gender and racial biases in both commercial and open-source AI generators, which are more pronounced than current societal disparities, but also more nuanced prejudices in facial expressions and appearances, which can subtly shape perceptions and reinforce stereotypes. As generative AI becomes increasingly ubiquitous, understanding and addressing these biases is essential to ensure the development of general-purpose technologies that truly benefit all members of society.

# 1. Introduction

Artificial intelligence (AI) is an indisputable catalyst of societal transformation, reshaping both our professional and personal landscapes. A pivotal player in this transformation is generative AI, with its potential to augment productivity and boost global economies. It is predicted that generative AI could add trillions of dollars in value to the global economy and significantly increase global GDP and productivity growth[1,2]. Generative AI's ability to generate a wide array of output forms—from text and code to images and videos—could potentially outpace human production capacity by 2030[3]. Particularly in marketing and sales, generative AI has the capacity to revolutionize the generation of creative content and reshape customer interactions, accounting for a substantial portion of the projected value of AI use cases[1]. Not surprisingly, Generative AI is already being used extensively in marketing for copywriting and creative generation[4]. The partnership between technology titan Nvidia and the world's largest advertising agency, WPP, further exemplifies the transformative potential of generative AI in marketing[5]. Given that marketing is dominant in shaping individuals' perceptions, we believe that generative AI will have a significant impact on shaping society.

Furthermore, the impact of AI extends beyond commercial sectors, penetrating the realm of education. AI's capacity to personalize learning experiences across all educational levels has been recognized[6]. Generative AI can improve the learning process by providing tailored information, stimulating creativity, and enhancing digital skills, preparing learners for future workplace demands. In the context of digital publishing in education, generative AI's potential for significant cost reduction of content creation (e.g., text, image, audio, video) underscores its practical benefits[7].

While generative AI has numerous benefits in business and society, we must also remain aware and vigilant for potential drawbacks. Concerns arise about the technology's potential risks, including issues related to intellectual property rights, accuracy of output, explainability of results, and potential propagation of harmful bias. Of significant concern is the potential bias embedded in generative AI models. Different from traditional AI models that are often used for classification or prediction, generative AI models are used to create new content based on patterns from the training data, making it difficult to measure the bias as there is no single "correct" output. Instead, one would need to evaluate a range of generated content for patterns that reflect bias. Moreover, new content generated by these models such as visual content can directly shape users' perceptions, perpetuate harmful stereotypes, and even distort their beliefs, especially if the generated content is widely disseminated. For example, as educational tools increasingly harness generative AI, they inherently mold young minds and shape their worldview. Content tailored to student characteristics with potential bias may inadvertently propagate or solidify harmful stereotypes, shaping perceptions in ways that may be challenging to reverse.

Since generative AI models are often trained on vast quantities of data collected from the internet, lack of control over the sources presents a formidable challenge in auditing and updating the training data to handle potential bias. As the training data may encompass a plethora of perspectives, cultures, and ideologies, it becomes increasingly challenging to anticipate, let alone correct, the myriad biases that could inadvertently find their way into the model. Furthermore, the situation is complicated by the proprietary nature of several major generative AI models. These models, often owned and maintained by private entities, are not publicly available for scrutiny. This lack of transparency further exacerbates the problem, as the broader academic community and society cannot easily assess or rectify bias in such closed systems. Given their increasing popularity and rapid adoption in many domains, generative AI models may inadvertently reinforce detrimental biases and stereotypes in our society, particularly in marketing and education. The current lack of well-defined regulations and policies in the domain of generative AI presents further risks, creating space for potentially harmful applications.

This study investigates potential bias in three of the most popular text-to-image AI generators—Midjourney, Stable Diffusion, and DALL·E 2. These AI tools, ranging from commercial to open-source, can generate high-quality images from text prompts, a capability used by millions of users and deployed across various domains. Fig. 1 presents some examples of images generated by Midjourney with different textual descriptions.

Prompt: A Portrait of Personal Financial Advisors

Prompt: A Portrait of Chief Executives

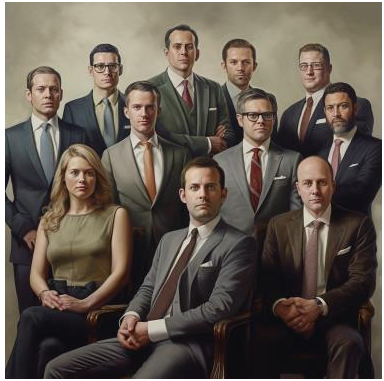
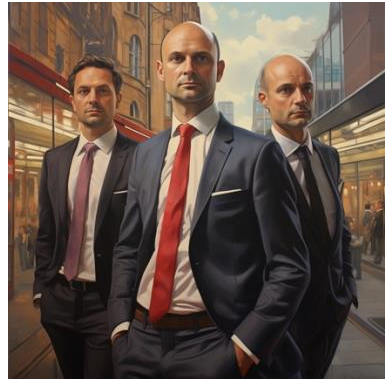

Prompt: A Portrait of Dishwashers

Prompt: A Portrait of Anesthesiologist Assistants

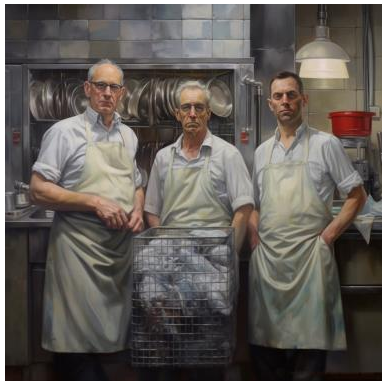
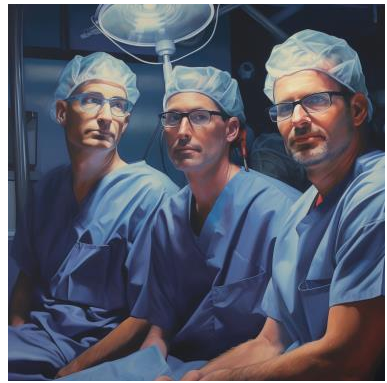

**Fig. 1. Examples of Images Generated by Midjourney with Different Prompts.**

Using approximately 8,000 occupational portraits produced by these three AI generators, we illuminate two primary areas of bias across all three generative AI models: (1) systematic gender and racial biases, and (2) subtle biases in facial expressions and appearances. Firstly, we find that women and Black individuals are significantly underrepresented in images generated by these tools. Alarmingly, when comparing this underrepresentation to different benchmarks such as BLS Labor Force Statistics[8] and Google images, the disparity is even more pronounced than the *status quo*, intensifying the biases and stereotypes we are actively striving to rectify in today's society. These biases may inadvertently perpetuate harmful societal stereotypes, with potentially far-reaching implications. For example, the overrepresentation of males in occupational images could deter aspiring female professionals, especially those at young ages when their minds are easily impressionable, thereby undermining efforts to promote Equity, Diversity, and Inclusion (EDI).

Secondly, beyond the blatant gender and racial disparities, our research highlights a more insidious form of bias in facial expressions and appearances. Women are often depicted as younger, more smiling and happier, while men are depicted as older, more neutral and angrier, and perhaps more authoritative. These nuanced biases, albeit subtle, might be even more treacherous. They can subconsciously embed and reinforce harmful gendered stereotypes, potentially conveying that women are more submissive or less competent than their male counterparts.

While the magnitude of bias varies among the tools, the direction of bias remains consistent across both commercial and open-source AI generators. As generative AI tools become increasingly used in various domains in society, it is imperative to rectify these biases before they cause harm at large scale. Ideally, generative AI should serve all of humanity equally, shaping an inclusive future free from harmful biases. Our study underlines the critical need to identify and mitigate these biases, reinforcing the commitment to ensuring that generative AI inclusively benefits all of humanity.

## 2. Research Context and Data Generation

We utilized three popular text-to-image AI generators – Midjourney, Stable Diffusion, and DALL·E 2 – to generate images for various occupations. Each generator employs distinctive models to transform text prompts into visual content. Notably, both Midjourney and DALL·E 2 are commercial AI image generators, while Stable Diffusion is an open-source program.

Midjourney was first published in July 2022 and is one of the leading drivers of emerging technology that uses AI to create visual imagery from text prompts[9], attracting approximately 15 million users[10]. Midjourney operates on a subscription-based model. DALL·E 2 is latest version of DALL·E, introduced by OpenAI in April 2022, and has amassed more than one million users[11]. Its services are offered on a pay-per-use basis. Stable Diffusion, released for free in August 2022, utilizes a latent diffusion model to generate high-quality images from textual descriptions. As an open-source tool, Stable Diffusion has been used in pop music videos, Hollywood movies, and is used by more than 10 million people daily[12]. The prowess of these generative AI models is well-recognized across diverse domains. For example, an AI-generated image was recently awarded the prize in an art competition, marking one of the first instances of AI-created artwork being so recognized[13]. This event signals a pivotal shift in societal and artistic perceptions and acceptance of AI-generated art.

To construct our data, we used consistent text prompts for all three models to generate images for occupations in the O*NET database, which contains information on 1,016 occupations[14]. It is a regularly updated database of occupational characteristics and worker requirements information across the U.S. economy, representing one of the world's largest, most comprehensive, widely used public repositories documenting detailed job-worker characteristics. Specifically, we structured our prompt as "A portrait of X" with X being each occupation in the O*NET database. For DALL·E 2, which incurs a cost for each generated image, we produced two images per prompt. Similarly, with Midjourney, we followed this process to create corresponding occupational portraits. Since Midjourney automatically returns four images for every prompt, we obtained four images for each occupation. As for Stable Diffusion, an open-source tool, we used a pre-trained model called Realistic Vision, which is one of the best Stable Diffusion models for generating realistic images. We used this model to generate two images for every prompt.

Consequently, about 8,000 different images were generated in total. We then conducted image analysis to investigate potential biases and stereotypes embedded in these AI-generated portrayals of various occupations.

## 3. Analyses and Results
### 3.1. Systematic Gender and Racial Biases

In order to determine the gender distribution within our data, we utilized the Face++ API, one of the best face recognition APIs on the market[15], to detect faces in each image and extract various features such as gender, smile, emotion, and age. Following this, we assessed the gender distribution within each image by calculating the percentage of women or men depicted, and subsequently computed an average across all images created by the same AI generators. To quantify gender and racial biases, we first define fairness based on statistical parity (also called demographic parity) which is one of the most popular fairness notions in machine learning. It simply requires that decisions or outcomes should be independent of sensitive attributes like gender or race, ensuring an equal distribution across demographic groups. Bias, therefore, is identified when there is a deviation from this statistical parity, manifesting as unequal representation among different groups. As summarized in Fig. 2, the results revealed that all three AI image generators exhibited bias against women. Specifically, the representation of women in images created by Midjourney, Stable Diffusion, and DALL·E 2 was 23%, 35%, and 42%, respectively, all significantly lower than their male counterparts, which were represented at 77%, 65%, and 58%, respectively ($p < 0.001$). While the degree of bias varied depending on the model, the direction of bias remained consistent across all three AI image generators, including both commercial and open-source models. Notably, about half of the occupational portraits generated by Midjourney and DALL·E 2 lacked any female representation as shown in Fig. S1.

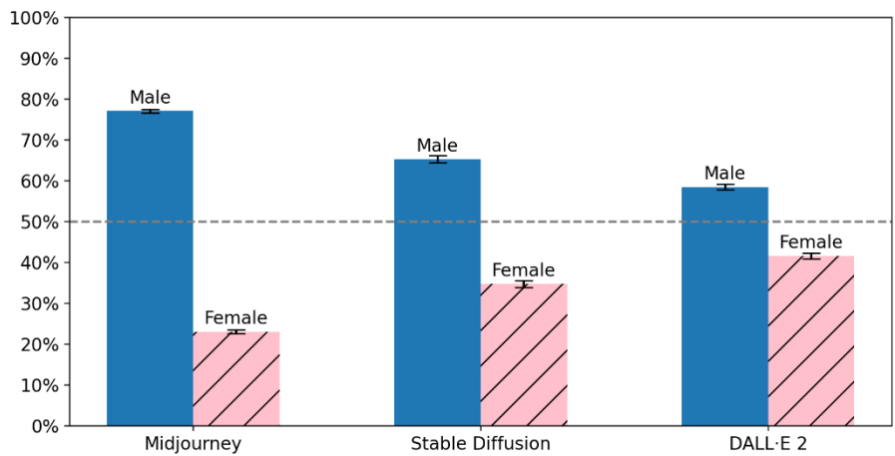

**Fig. 2. Gender Distribution in Occupational Portraits Created by AI Generators**

While the disparity in the generated images is alarming at an aggregate level, we examine how this disparity manifests across different types of jobs. To do this, we utilized a descriptor from the O*NET database known as "Job Zone" as a proxy. A job zone categorizes occupations that require similar levels of education, related experience, and on-the-job training. Within the O*NET database, there are 5 job zones, with Job Zone 1 requiring minimal preparation (3 months), and Job Zone 5 necessitating the most extensive preparation, typically over 4 years. Table S1 outlines the differences across job zones[16]. It is observed that as the level of preparation required increases, so does the median income, with workers in Job Zone 1 earning $30,230, and those in Job Zone 5 earning $81,980.

We then calculated the gender distribution in the portraits of occupations within each job zone. The findings, summarized in Fig. 3 (A), indicate that gender bias permeates all job zones within the images generated by Midjourney and Stable Diffusion, with the extent of bias slightly decreasing as the required level of preparation increases. For DALL·E 2, the gender bias is evident across the first four job zones, with the representation of females surpassing that of males in Job Zone 5, which necessitates the highest level of preparation. A potential explanation for these findings is that jobs requiring lower levels of preparation might have exhibited stronger gendered representations historically. Such historical bias could be reflected in the data the models were trained on, leading the AI to generate images that not only mirror but potentially amplify these long-standing stereotypes. For example, recent data[8] indicates an increasing presence of female workers across job zones: 39% in Job Zone 1, 38% in Job Zone 2, 43% in both Job Zone 3 and 4, and 59% in Job Zone 5.

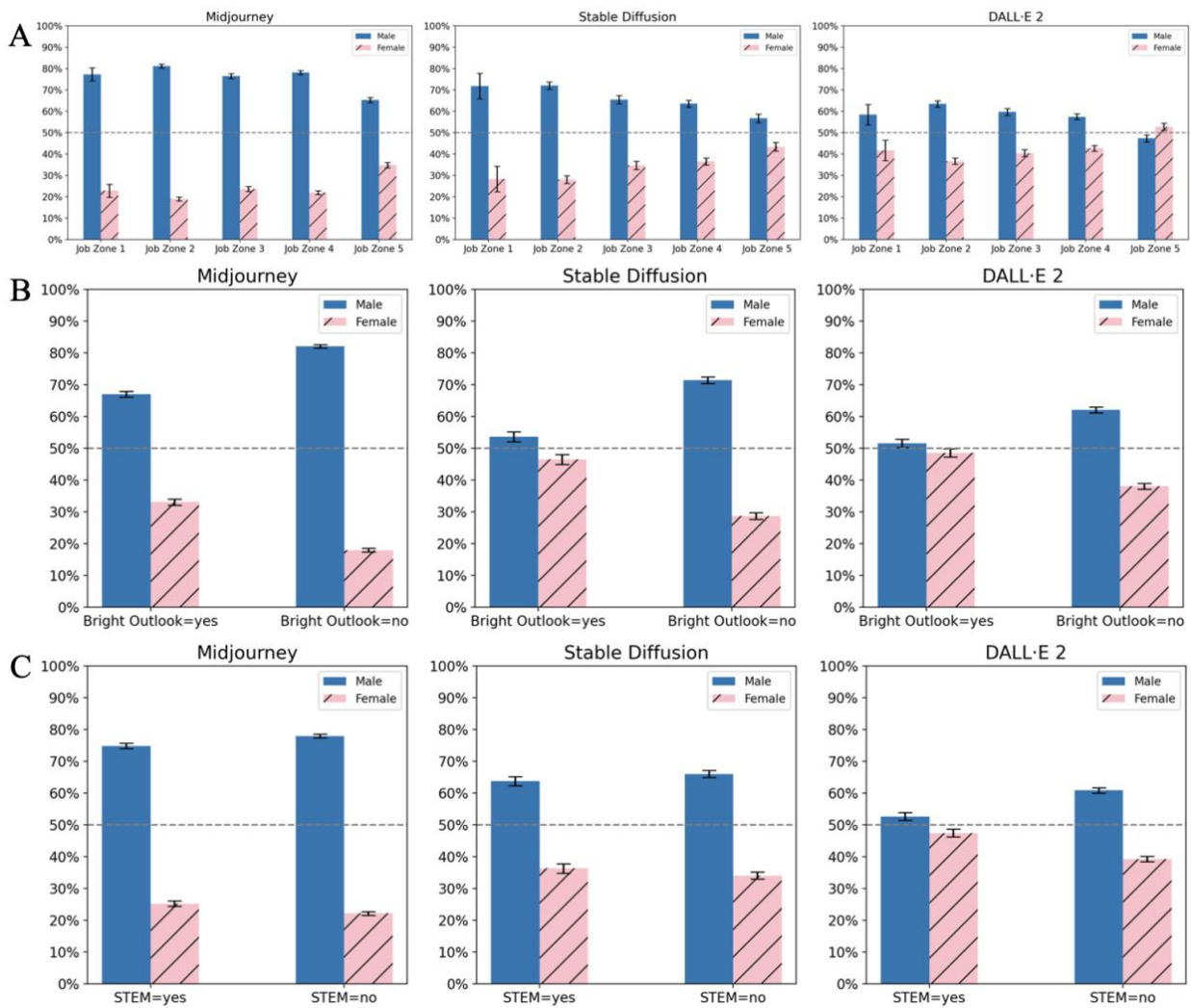

**Fig. 3. Gender Distribution in Portraits of Different Occupations Created by AI Generators**

Next, we investigate the variance in gender bias in relation to types of occupations using another proxy, an O*NET occupation-level indicator called the "Bright Outlook." Bright Outlook occupations are those expected to grow rapidly in the next several years, have large numbers of job openings, or are new and emerging occupations. Of the 1,016 occupations in the O*NET database, 340 are classified as "Bright Outlook." This indicator is positively correlated with the job zones (Pearson correlation coefficient = 0.131, $p<0.001$). We calculated the gender distribution in the portraits of Bright Outlook occupations, the results of which are depicted in Fig. 3 (B). The gender bias was evident in all types of occupations across the three AI image generators, with a reduced bias in the portraits of Bright Outlook occupations.

To further investigate the variance in gender bias according to types of occupations, we used another O*NET occupation-level indicator, "STEM," denoting occupations that require education in science, technology, engineering, and mathematics (STEM) disciplines. Among the 1,016 occupations in the O*NET database, 287 are labeled as "STEM." This indicator is also positively correlated with the job zones (Pearson correlation coefficient = 0.563, $p<0.001$). We calculated the gender distribution in the portraits of STEM occupations, the results of which are summarized in Fig. 3 (C). While the gender bias was still prevalent across all types of occupations for the three AI generators, it was less pronounced in the portraits of STEM occupations.

Next, we evaluated racial distribution by measuring the percentage of Black and White individuals portrayed in images produced by the three AI models. Utilizing the DeepFace framework[17], we categorized the race of each face in the images as either White, Black, Asian, Latino, Indian, or Middle Eastern to compute the racial distribution in each image. We subsequently calculated the average value across all images produced by the same AI generators. Our findings, summarized in Fig. 4, reveal that Black individuals are less likely to be generated than White individuals across all three AI generators ($p < 0.001$). Specifically, the percentage of Black individuals in the occupational portraits generated by Midjourney, Stable Diffusion, and DALL·E 2 was 9%, 5%, and 2%, respectively. For Stable Diffusion, however, the representation of Asians exceeded 50%. This disproportion might be partially attributed to the training data used for the pre-trained Stable Diffusion model. Notably, for the two commercial AI generators, Midjourney and DALL·E 2, the representation of White individuals exceeded 50% consistently across all five job zones as shown in Fig. S1.

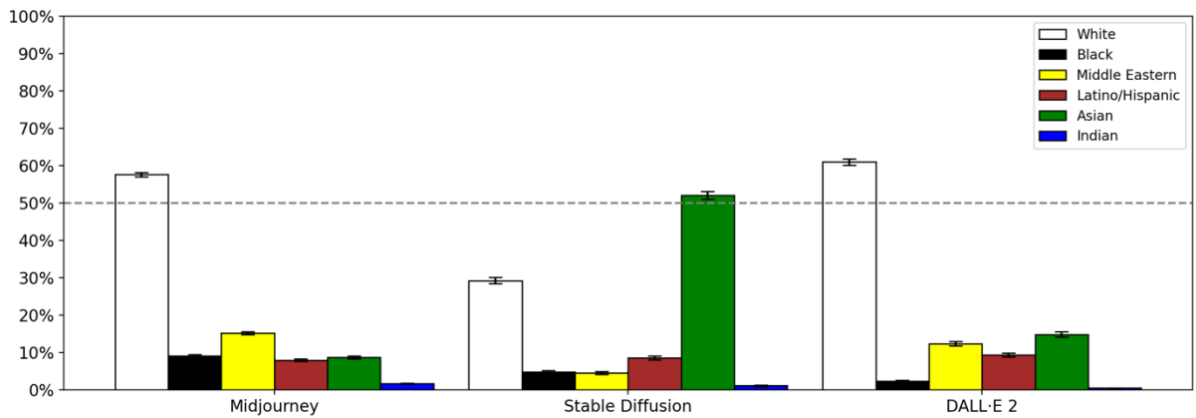

**Fig. 4. Average Racial Distribution in Images Created by AI Generators**

The gender and racial bias in the portraits of occupations generated by AI tools may stem from the inherent bias present in the training data, which reflects real-world disparities. To examine this, we obtained employment and demographic data from the BLS Labor Force Statistics[8], from the Current Population Survey collected by the U.S. government, primarily capturing non-self-employed, documented workers in the formal economy. This data reported that the percentage of female workers of total employed is 46.8%[8], significantly higher than the percentage of females in portraits of occupations created by Midjourney ($p < 0.001$), Stable Diffusion ($p < 0.001$), and DALL·E 2 ($p < 0.05$); the percentage of Black workers is 12.6%[8], significantly higher than that in occupational portraits created by Midjourney ($p < 0.05$), Stable Diffusion ($p < 0.001$), and DALL·E 2 ($p < 0.001$). A comparison of their distribution is presented in Fig. 5. The results show that the gender and racial disparities in the real world are amplified in all three AI generative models. This is especially concerning given that the benchmark itself (i.e., employment and demographic data from the BLS Labor Force Statistics) might encapsulate the societal bias that we are striving to counteract through various EDI initiatives.

Furthermore, we examined occupation-specific disparities to pinpoint where the bias was most evident. Specifically, for each dataset, we calculated the disparities in female and Black representation relative to the BLS benchmarks. Occupations were then ranked based on these disparities. The results, presented in Fig. S1-S6 in the supplementary materials, reveal pervasive gender and racial bias across most occupations. Interestingly, results from the three models show some consistent patterns. For example, "food preparation and serving related workers" consistently appear among the top three occupations exhibiting the most pronounced gender bias against women across all datasets, and "postal service clerks" rank at the top for the most pronounced racial bias against Black individuals in all three models.

To assess how the generative AI models compare to publicly available data, we collected images from Google Image Search as another benchmark. To construct this benchmark, we used the same text prompts as search keywords in Google to collect images representing different occupations in the O*NET database. For each search keyword, we saved the first ten images returned in the search results using Google's API. This resulted in 10,160 images in total. These images are then analyzed to measure the gender and racial distribution in the data. Our results show that the percentage of females in the data collected from Google is 44.5%, which is not statistically different from that of BLS ($p > 0.5$), but significantly higher than the percentage of females in portraits of occupations created by Midjourney ($p < 0.001$), Stable Diffusion ($p < 0.001$), and DALL·E 2 ($p < 0.05$). This indicates that the image data on Google is representative of the gender distribution in the population and is perhaps not contributing to the gender bias in image generation models as far as gender is concerned.

The percentage of Black individuals in this Google data is 5.41%, which is significantly lower than that of BLS and Midjourney ($p < 0.001$), not statistically different from that of Stable Diffusion, but significantly higher than the percentage of Black individuals in portraits of occupations created by DALL·E 2 ($p < 0.001$). A comparison of the distribution is illustrated in Fig. 10, showing that all three AI generative models exhibited amplified gender and racial disparities in their generated images, which is alarming given that both benchmarks themselves may encapsulate the societal bias that we are trying to rectify through various EDI initiatives nowadays.

We argue that such gender and racial biases in representation are problematic, especially when these biases are amplified beyond real-world disparities and are worse than the status quo. For example, portraying primarily men may potentially dissuade the next generation of female and Black professionals and hinder efforts to promote EDI. Generative AI should benefit all of humanity and be shaped to be as inclusive as possible, at least not amplifying the biases in the status quo. Rather than reflecting, or even amplifying, the existing biases of today's world, these tools should aspire to shape a better future that reflects equality and fairness.

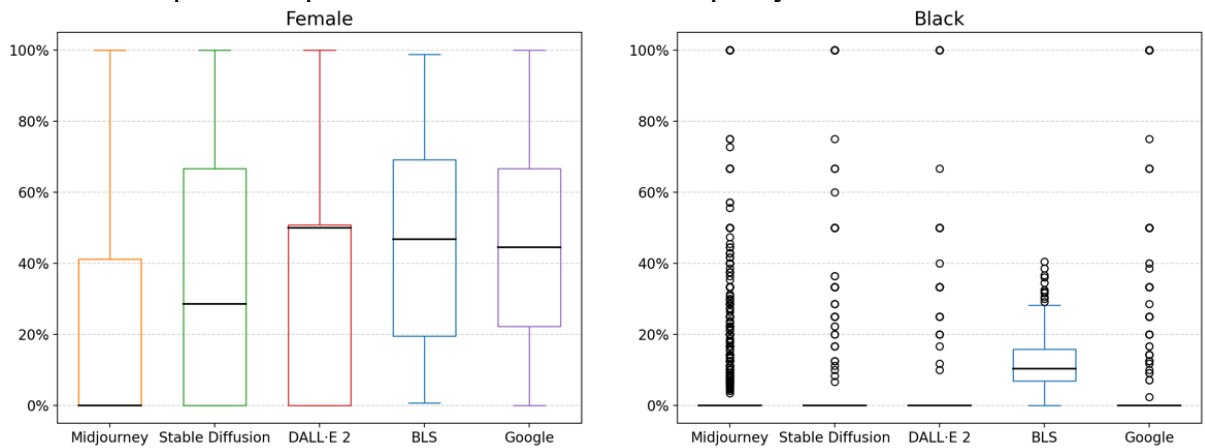

**Fig. 5. Comparing Gender Distribution Across Datasets**

## *3.2. Subtle Biases in Facial Expressions and Appearances*

It is important not only to consider who is shown in these images, but in which light they are shown. Facial expressions, age, approachability, and authority are various dimensions of these images that can impact how consumers perceive them. A recent paper shows that individuals with expressions of anger and disgust are judged to be more dominant than those expressing happiness[18]. Drawing upon features such as gender, age, smile, and emotions extracted from detected faces, we gauged the average facial expressions and appearances for men and women in each image, and the outcomes are represented across the image generations. These results are illustrated in Fig. 6, suggesting that men appear older and more senior than women in images generated by all three AI models which might inadvertently portray them in a position of authority as opposed to women that typically exhibit more smiles in those images. Moreover, women express more happiness than men across all three AI generators. Men, on the other hand, typically express more neutral emotion and anger than women across all three AI generators. This stereotype persists across all three generative models.

We argue that such consistent disparities in representation is problematic. It is worth noting that smiling has historically been interpreted as a submissive gesture conveying deference or appeasement[19,20]. Smiling faces have been perceived to be less dominant and more submissive than nonsmiling faces[21]. Past studies suggest that individuals exhibiting dominance are often perceived as high in status[22] and more competent[23,24]. Moreover, prior research has shown that both perceived dominance and perceived affiliativeness are significantly related to perceptions of emotionality[25]. Specifically, perceived dominance is significantly positively related to the disposition to show anger, disgust, and contempt, whereas perceived affiliativeness is significantly positively related to the perceived disposition to show happiness, surprise, and fear[25]. Given these historical and cultural connotations, there is a risk that these generative AI models unintentionally depict women as more submissive and less competent than men.

Such nuanced biases, subtle yet consistent across all three generative AI tools, can be insidious in their impact, weaving their way into societal perceptions without immediate notice. Over time, they can contribute to shaping collective unconscious biases that further perpetuate gender inequalities, impacting everything from workplace dynamics to societal norms. Moreover, these subtler biases present unique challenges in rectification, as they might not be immediately recognized as problematic due to their subtle nature. Addressing them requires not just technical solutions but a deeper cultural and psychological understanding of biases and their origins. The potential harms extend beyond mere perceptions. For example, if such AI tools are used in educational settings, young learners might internalize these biased representations, shaping their aspirations and self-perceptions. In corporate settings, these portrayals might influence hiring decisions, performance evaluations, and even team dynamics, amplifying the algorithmic and social biases even further.

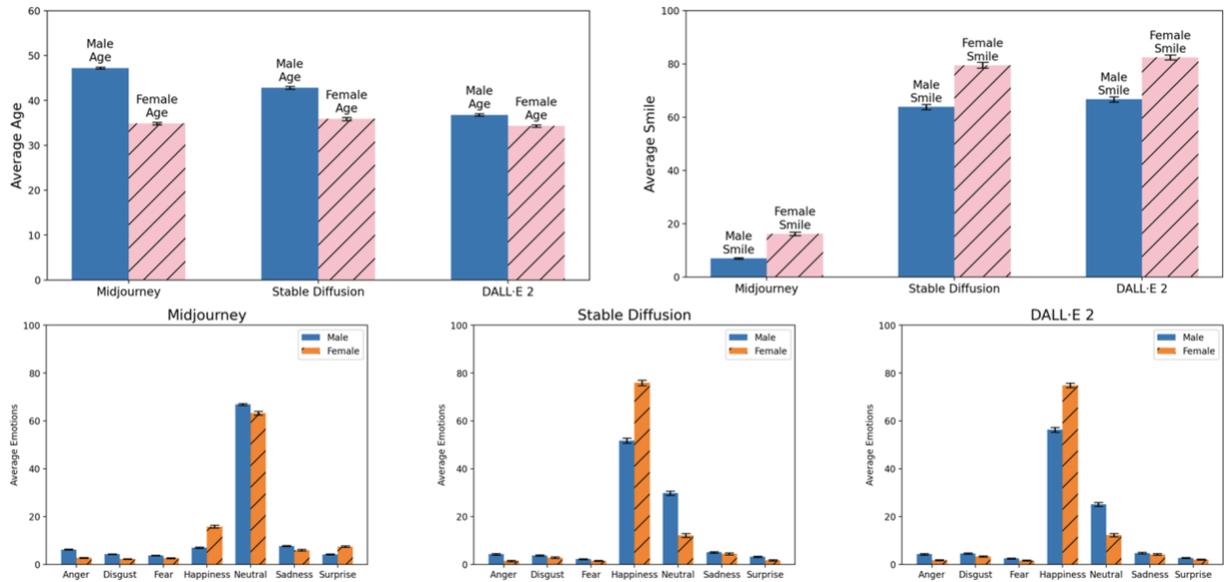

**Fig. 6. Average Age, Smile, and Emotions of Males vs Females in Images Generated by AI**

## 4. Conclusions

This study highlights the presence of two principal biases embedded within AI image generators: systematic gender and racial biases and subtle biases in facial expressions and appearances. Our research underscores the urgency of addressing these issues, as these AI tools are increasingly employed in diverse sectors including marketing and education. The potential fallout of such biases, including the shaping and reinforcement of societal perceptions and stereotypes, warrants our meticulous attention and exploration.

By illuminating how AI interprets different occupations, our research contributes to a crucial understanding of the bias endemic in AI systems. The degree of bias pervasive within these generative AI tools is a matter of significant concern, calling for immediate introspection and remedial action within the AI community.

As our reliance on AI intensifies, it becomes imperative to ensure fairness and equity in the design and deployment of these technologies. We must prioritize the development of generative AI systems that are not only technologically advanced but are also shaped by an ethical commitment to inclusivity and equity. This ensures these powerful tools are designed to benefit all users equitably, fostering a more inclusive future.

15. Eden AI, Best face recognition APIs in 2023 (2023); https://www.edenai.co/post/best-face-recognition-apis.

16. T. Eloundou, S. Manning, P. Mishkin, D. Rock, GPTs are GPTs: An early look at the labor market impact potential of large language models. arXiv:2303.10130 (2023).

17. S. I. Serengil, A. Ozpinar, Lightface: A hybrid deep face recognition framework. In *2020 Innovations in Intelligent Systems and Applications Conference* (ASYU) (IEEE, 2020), pp. 1-5.

18. Y. Ueda, S. Yoshikawa, Beyond personality traits: Which facial expressions imply dominance in two-person interaction scenes?. *Emotion* (2018), 18(6), 872.

19. E. Goffman, *Gender advertisements* (1979). New York: Harper & Row.

20. N. M. Henley, *Body politics: Power, sex, and nonverbal communication* (1977). Englewood Cliffs, NJ: Prentice-Hall.

21. C. F. Keating, A. Mazur, M. H. Segal, P. G. Cysneiros, W. T. Divale, J. E. Kilbride, S. Komin, P. Leahy, B. Thurman, R. Wirsing, Culture and the perception of social dominance from facial expressions. *Journal of Personality and Social Psychology* (1981), 40(4), 615-626.

22. C. Ridgeway, Nonverbal behavior, dominance, and the basis of status in task groups. *American Sociological Review* (1987), 52, 683-694.

23. J. S. Wiggins, R. Broughton, The interpersonal circle: A structural model for the integration of personality research. *Perspectives in Personality* (1985), 1, 1-47.

24. J. S. Wiggins, N. Phillips, P. Trapnell, Circular reasoning about interpersonal behavior: Evidence concerning some untested assumptions underlying diagnostic classification. *Journal of Personality and Social Psychology* (1989), 56, 296-305.

25. U. Hess, R. Adams Jr, R. Kleck, Who may frown and who should smile? Dominance, affiliation, and the display of happiness and anger. *Cognition and Emotion* (2005), 19(4), 515-536.
12

# Supporting Information

Table. S1. Occupations by Job Zone

| Job Zone | Preparation Required | Education Required | Example Occupations | Median Income |
|---|---|---|---|---|
| 1 | None or little (0-3 months) | High school diploma or GED (optional) | Food preparation workers, dishwashers, floor sanders | $30,230 |
| 2 | Some (3-12 months) | High school diploma | Orderlies, customer service representatives, tellers | $38,215 |
| 3 | Medium (1-2 years) | Vocational school, on-the-job training, or associate's degree | Electricians, barbers, medical assistants | $54,815 |
| 4 | Considerable (2-4 years) | Bachelor's degree | Database administrators, graphic designers, cost estimators | $77,345 |
| 5 | Extensive (4+ years) | Master's degree or higher | Pharmacists, lawyers, astronomers | $81,980 |

Fig. S1. Percentage of Different Images Created by AI Generators

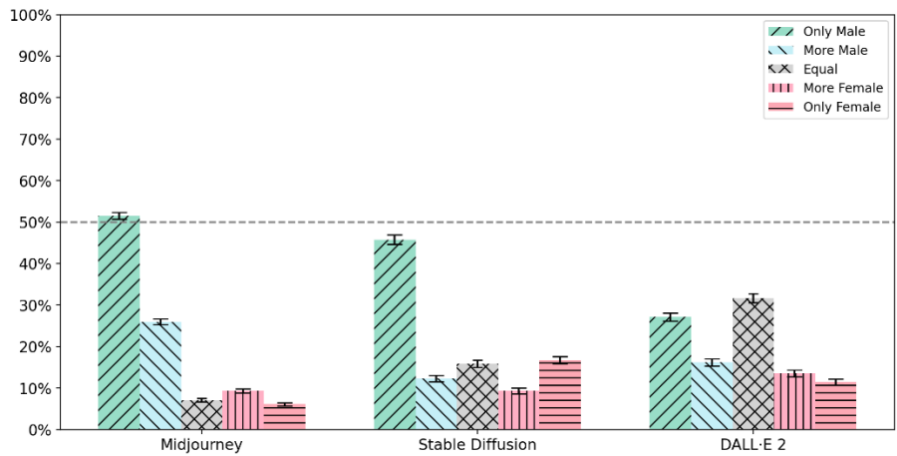

We classified each image into one of five exclusive categories: "Only Male" if it solely depicted males, "More Male" if the number of males exceeded females, "Equal" if there was an equal number of males and females, "More Female" if the females outnumbered males, and "Only Female" if it exclusively portrayed females. As summarized in Fig. S1, there was a significant skew towards male representation across all three AI generators. Notably, 52% of the images generated by Midjourney and 46% of the images generated by Stable Diffusion contained only males, indicating that approximately half of the generated occupational portraits by these models lacked any female representation.

Fig. S2. Average Racial Distribution by Job Zone in Images Created by AI Generators

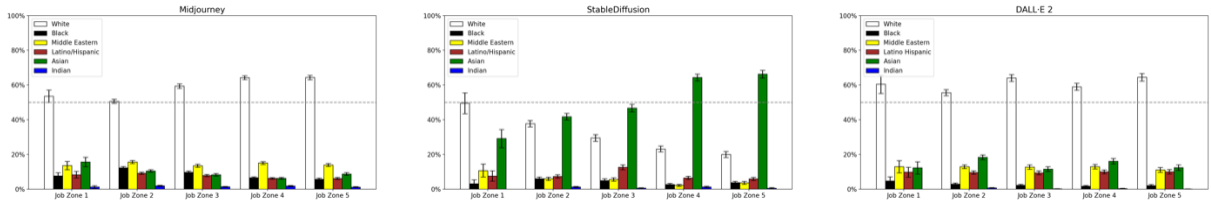

As shown in Fig. S2, for the two commercial AI generators, Midjourney and DALL·E 2, the representation of White individuals exceeded 50% consistently across all five job zones. However, the gaps between different races do not decrease as the required level of preparation increases, which was observed in the gender analysis. This is likely because the historical bias against Black individuals does not diminish across job zones. For example, recent data (BLS 2022) shows a fluctuating presence of Black workers across job zones: 13% in Job Zone 1, 16% in Job Zone 2, 11% in Job Zone 3, and 10% in both Job Zone 4 and 5.

Fig. S3. Disparity in Female Ratios: Midjourney vs BLS Data by Occupation

Fig. S4. Disparity in Female Ratios: Stable Diffusion vs BLS Data by Occupation

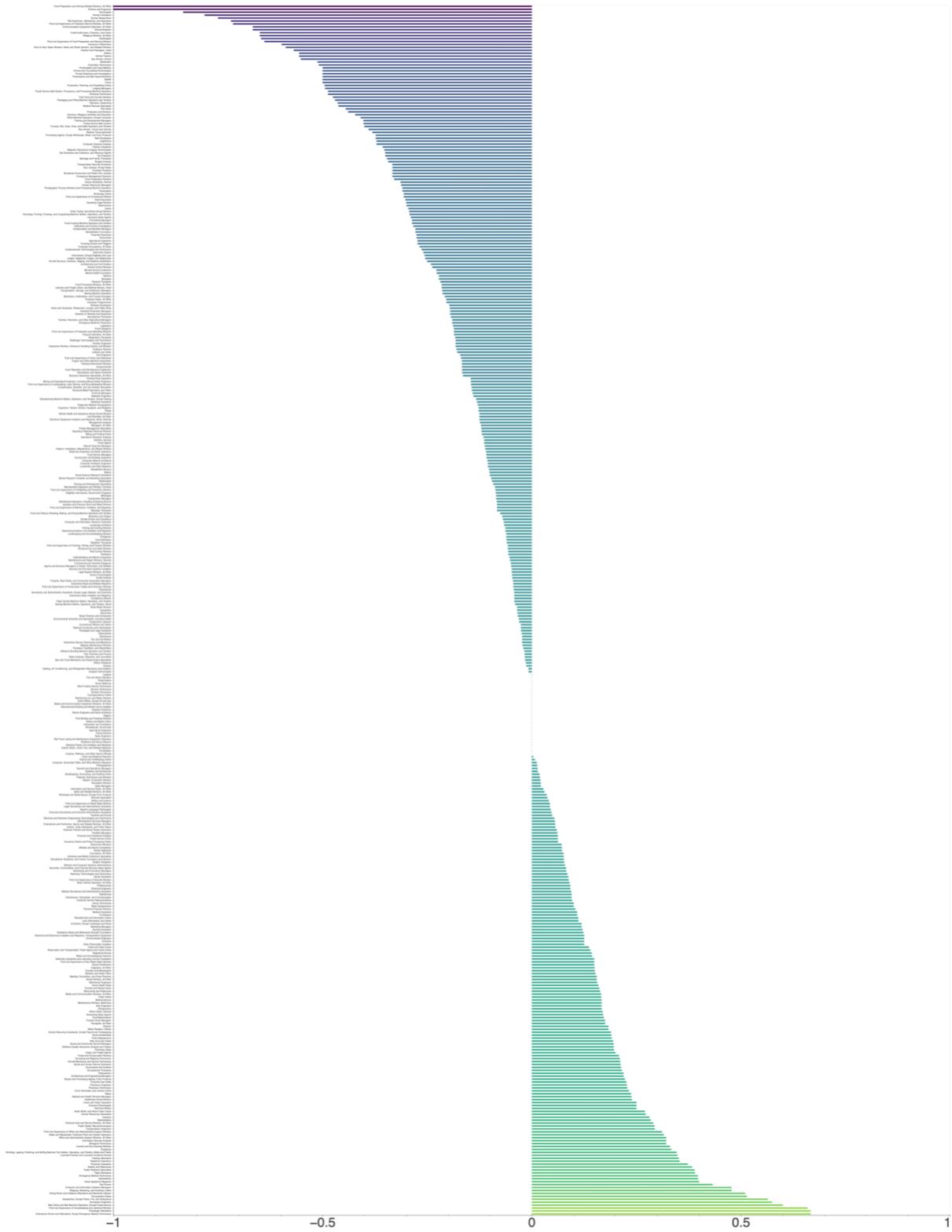

Fig. S5. Disparity in Female Ratios: DALL·E 2 vs BLS Data by Occupation

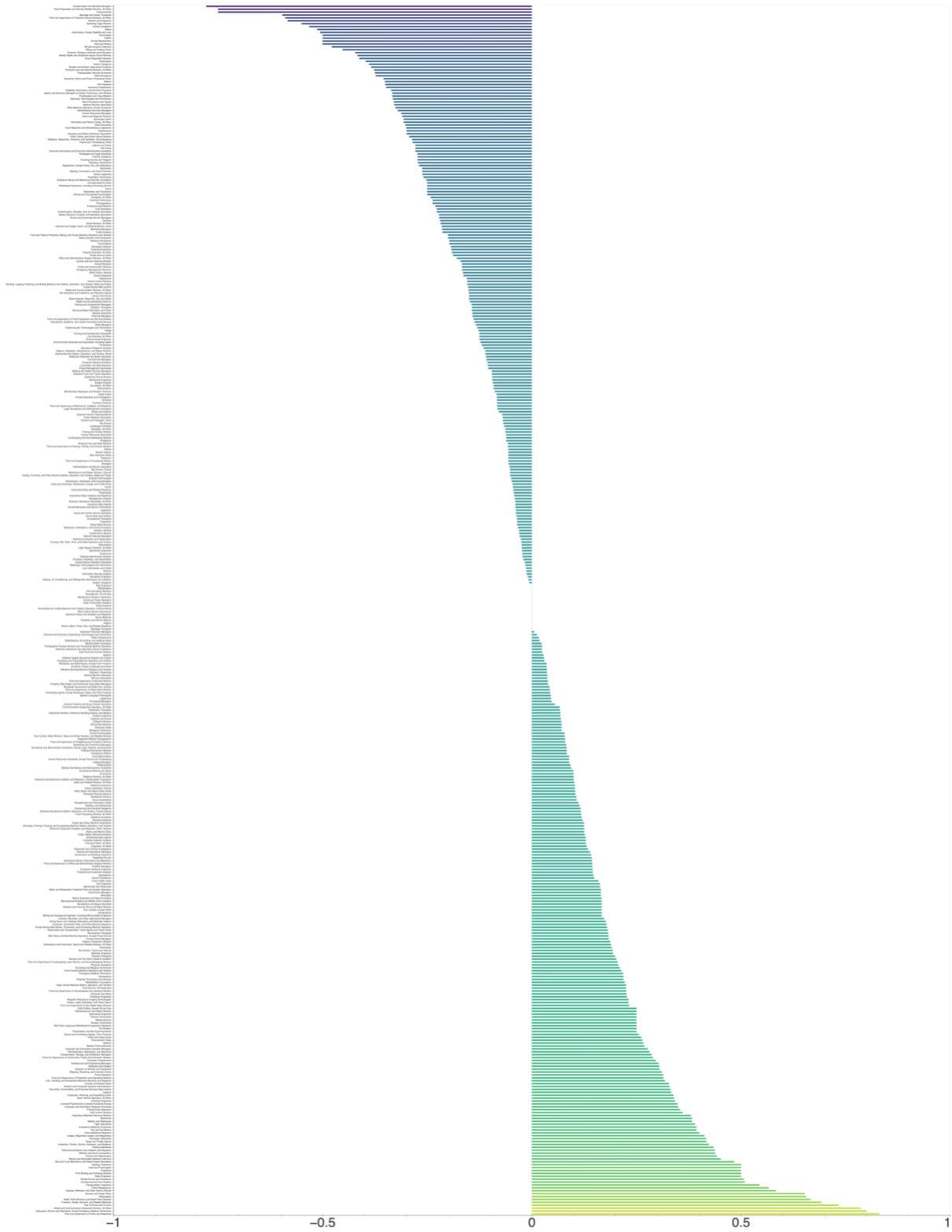

Fig. S6. Disparity in Black Ratios: Midjourney vs BLS Data by Occupation

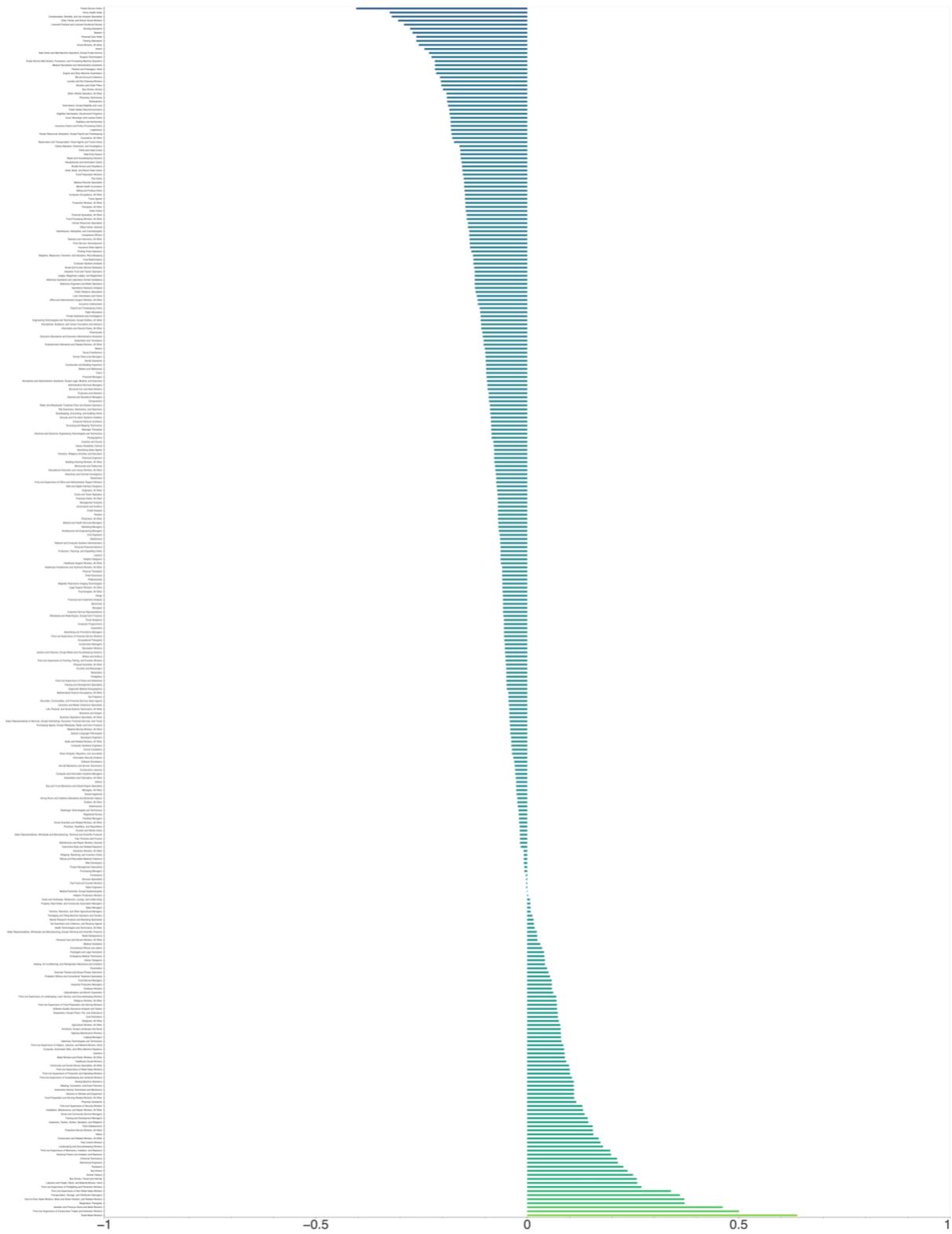

Fig. S7. Disparity in Black Ratios: Stable Diffusion vs BLS Data by Occupation

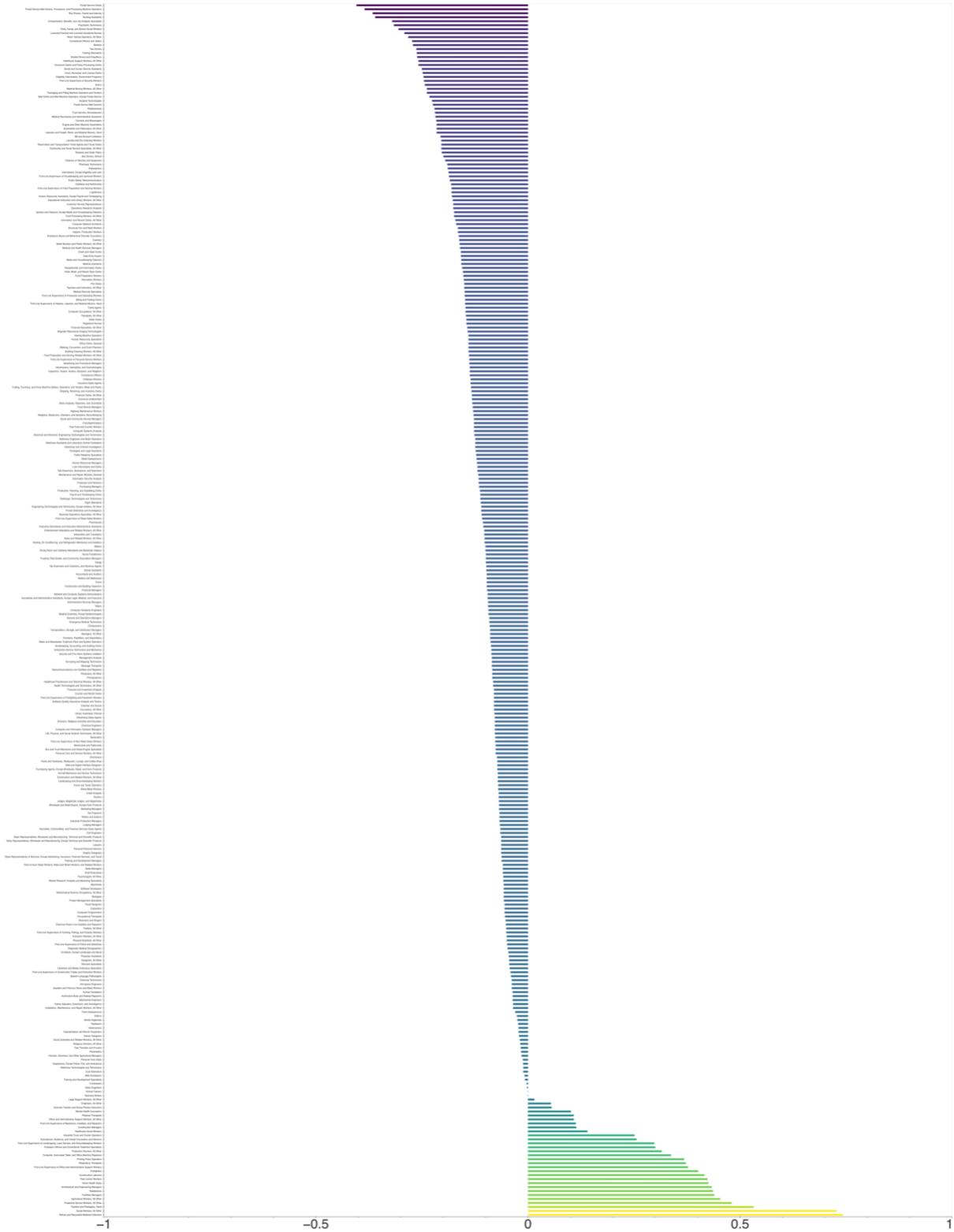

Fig. S8. Disparity in Black Ratios: DALL·E 2 vs BLS Data by Occupation

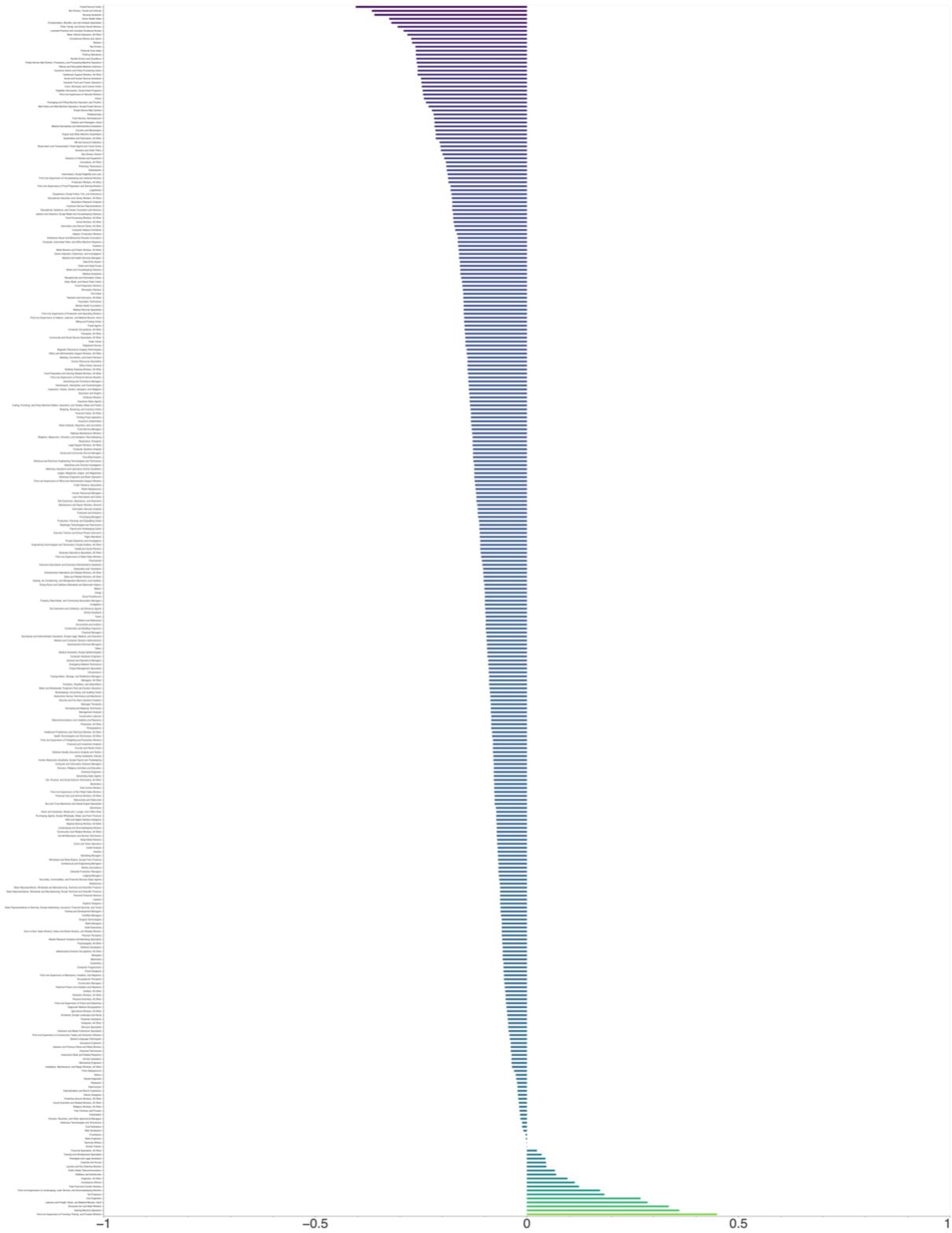